\documentclass[aps,prl,twocolumn,showpacs,floatfix]{revtex4}
\usepackage{graphicx}
\usepackage{bm}
\usepackage{amsmath}

\begin{document}
\title{Synchronization of extended chaotic systems with long-range
  interactions:\\ an analogy to L{\'e}vy-flight spreading of
  epidemics} 

\author{Claudio Juan Tessone$^{1,2}$, Massimo
  Cencini$^{3}$, and Alessandro Torcini$^{2,4}$}

\affiliation{$^{1}$ Institut Mediterrani d'Estudis Avan\c{c}ats,
  CSIC-UIB, Ed.~Mateu Orfila, Campus UIB, 07122 Palma de Mallorca,
  Spain \\ $^{2}$ Istituto dei Sistemi Complessi - CNR, via Madonna
  del Piano 10, 50019 Sesto Fiorentino, Italy\\ $^{3}$ INFM-CNR, SMC
  Dipartimento di Fisica, Universit\`a di Roma ``La Sapienza'', p.zle
  A.\ Moro 2,\\ 00185 Roma, Italy and Istituto dei Sistemi Complessi -
  CNR, via dei Taurini 19, 00185 Roma, Italy\\ $^{4}$ INFN - Sezione
  di Firenze, via Sansone 1, 50019 Sesto Fiorentino, Italy}
\date{\today}

\begin{abstract}
Spatially extended chaotic systems with power-law decaying
interactions are considered. Two coupled replicas of such systems
synchronize to a common spatio-temporal chaotic state above a certain
coupling strength. The synchronization transition is studied as a
nonequilibrium phase transition and its critical properties are analyzed
at varying the interaction range. The transition is found to be
always continuous, while the critical indexes vary with continuity
with the power law exponent characterizing the interaction.  Strong
numerical evidences indicate that the transition belongs to the {\it
  anomalous directed percolation} family of universality classes found
for L{\'e}vy-flight spreading of epidemic processes.
\end{abstract}

\maketitle

Since its discovery~\cite{fujisaka83}, chaotic synchronization has
become a very active and important field of research~\cite{PikoBook},
especially for its applications in control and secure
communications~\cite{first,roy_05}. It has been observed in various
experimental setups from semiconductor lasers to chemical
reactions~\cite{roy01}. The synchronization transition (ST) is
particularly interesting in spatially extended chaotic systems
\cite{stc1} where, due to the spatial dependence, a parallel with
nonequilibrium phase transitions can be drawn.  Extensive numerical
studies of coupled map lattices (CML), which are prototype models for
space-time chaos~\cite{stc}, have shown that ST between
two replicas of CML with short range interactions is continuous and
the associated critical properties belong to two distinct universality
classes depending on the nature of the individual elements of the
system~\cite{baroni,PA01,munoz,GLPT03}.  For smooth chaotic maps,
linear mechanisms rule information propagation, and ST occurs as soon
as the transverse Lyapunov exponent crosses zero. In this case, the
transition belongs to the multiplicative noise universality
class~\cite{munoz_review}. For discontinuous maps, a linear
stability analysis is no more sufficient to locate the ST, and the
replicas synchronize (for definitely negative transverse Lyapunov
exponent) when the spreading velocity of finite amplitude perturbations
vanishes~\cite{baroni}. The synchronization in such systems is 
controlled by nonlinear effects and the transition belongs to the
directed percolation (DP) universality class~\cite{haye}, analogously to
what found for cellular automata~\cite{G99}.  At present we thus
have a satisfactory understanding of the critical properties of ST for
short-range coupled chaotic systems~\cite{PA01,munoz}.

The aim of this Letter is to study the synchronization transition of
extended chaotic systems with long range spatial coupling. Long range
interactions are indeed relevant to many real systems such as neuron
populations \cite{neurosynch}, Josephson junctions \cite{jose},
cardiac pacemaker cells \cite{cuore}, and to the issue of disease
spread triggered by aviation traffic~\cite{geisel}. We consider, as a
prototype of systems with long range interacting elements, a model of
CMLs with coupling decaying as a
power-law~\cite{PV94,anteneodo,cencini_torcini}. As a first step we
investigate the case of maps whose dynamics is dominated by nonlinear
mechanisms.  We show that the critical indexes characterizing the
transition depend continuously upon the exponent $\sigma$, which sets
the coupling range. Moreover, strong numerical evidences indicate that
the critical exponents are compatible with those found in a model for
\textit{anomalous directed percolation} studied by Hinrichsen and
Howard~\cite{HH98} (HH in the following).

Anomalous DP arises in epidemic spreading whenever the infective agent
can perform unrestricted L{\'e}vy-flights \cite{HH98,janssen}.  Such
processes, originally introduced in Ref.~\cite{mollison77}, can be
modeled by assuming, e.g. in $d\!=\!1$, that an already infected site
propagates the disease to any other site with a probability $P(r) \sim
r^{-(1+\sigma)}$ algebraically decaying with the distance $r$
(similarly to the distance distribution of human travels, which are
nowadays responsible for the geographical spread of infectious
diseases~\cite{geisel}).  Numerical studies of a stochastic lattice
model, which generalizes directed bond percolation, have shown that
the critical exponents vary continuously with $\sigma$~\cite{HH98},
despite the divergence of the average interaction distance for
$\sigma<1$ and of the second moment of $P(r)$ for $\sigma<2$. These
findings confirm previous theoretical results~\cite{janssen} according
to which usual DP should be recovered for sufficiently short-range
coupling (namely for $\sigma > \sigma_c \equiv 2.0677(2)$) and the
mean-field description should become exact for $\sigma < \sigma_m
\equiv 0.5$. Furthermore, close to $\sigma_m$, the critical exponents
found by HH agree with renormalization group
calculations~\cite{janssen}.

We investigate the synchronization transition of
two coupled replicas of CML defined as:
\begin{eqnarray}
x_i(t+1) &=& (1-\gamma) F(\tilde{x_i}(t))+\gamma F(\tilde{y_i}(t))\nonumber\\
y_i(t+1) &=& (1-\gamma) F(\tilde{y_i}(t))+\gamma F(\tilde{x_i}(t))
\label{eq:cml}
\end{eqnarray}
where $\tilde{z_i}=\{\tilde{x_i}, \tilde{y_i}\}$ is given by:
\begin{equation}
\tilde{z_i}=(1-\epsilon) z_i + \frac{\epsilon}{\eta(\sigma)}
\sum_{m=1}^{M} \frac{z_{i-j_m(q)}+z_{i+j_m(q)}}{(j_m(q))^\sigma}\,.
\label{eq:smartcoupling}
\end{equation}
The labels $t$ and $i$ are the discrete temporal and spatial indexes,
$x_i(t),y_i(t) \in [0:1]$ are the state variables, $L$ is the lattice
size ($i\!=\!1,\dots,L$), and periodic boundary conditions are assumed
(i.e. $z_{i+L}(t)=z_i(t)$).  As local dynamics, throughout this work,
we employ the Bernoulli map $F(x) \!=\! 2 x$ ({\rm mod} 1), which is
characterized by a discontinuity that makes the nonlinear effects
dominating~\cite{nota1}.  The interaction range is controlled by
the exponent $\sigma$: for $\sigma \!=\! 0$ globally coupled maps are
obtained, while standard CML's with nearest neighbor coupling are
recovered in the limit $\sigma \to \infty$.  
Two coupling constants $\epsilon$ and $\gamma$ enter the system
definition.  The former measures the amplitude of the spatial coupling
and is fixed to $2/3$.  The latter controls the strength of the
site-wise interaction between the two replicas. In the range of
$\sigma$ examined, a synchronized state is always achieved as the
coupling exceeds a certain $\gamma_c\!=\!\gamma_c(q,\sigma)$.  

Usually the model has been studied in the fully coupled
case~\cite{PV94,anteneodo,cencini_torcini}, i.e. for $j_m(q)\equiv m$
with $M\!=\!(L\!-\!1)/2$. Instead by following Ref.~\cite{told}, we consider
here a modified version of the coupling, where $j_m(q)=q^m -1$ with
$q$-values typically chosen as $q=2,4$ and $8$. The sum in
Eq.~(\ref{eq:smartcoupling}) extends to $M\!=\!\log_q (L/2)$; finally
$\eta(\sigma)\!=\!2 \sum_{m=1,M} (j_m(q))^{-\sigma}$ is a normalization
factor.  The rationale for the choice $j_m(q)\!=\!q^m \!-\!1$ lies in its
convenience from a computational point of view. As a matter of fact,
the chosen coupling scheme allows for simulating very large systems:
each updating step can be performed in $\mathcal{O}(L\log_q L)$
operations instead of $\mathcal{O}(L^2)$ needed in the fully coupled
case. Moreover, a simple reasoning shows that the model with
$j_m(q)\!=\!q^m\! -\!1$ and exponent $\sigma$ exhibits the same critical
properties of the fully coupled one with a different exponent
$\sigma_{fc} = \sigma+1$.  Indeed both versions of the model should
display the same critical properties once the spatial interactions
scale analogously. For the modified model the coupling weight over the
interval $]j_m(q),j_{m+1}(q)]$ containing a single coupled site is
    simply $1/j_{m+1}(q)^\sigma \sim q^{-(m+1)\sigma}$, while for the
    fully coupled case this amounts to
    $\sum_{k=j_m(q)+1}^{j_{m+1}(q)}{k^{-\sigma_{fc}}} \sim
    q^{-(m+1)(\sigma_{fc}-1)}$. Therefore, the two weights scale in
    the same manner only for $\sigma_{fc}\! =\! \sigma\! +\!1$ as confirmed by
    numerical results reported in the following.

\begin{figure}[t!]
\includegraphics[draft=false,clip=true,height=0.34\textwidth]{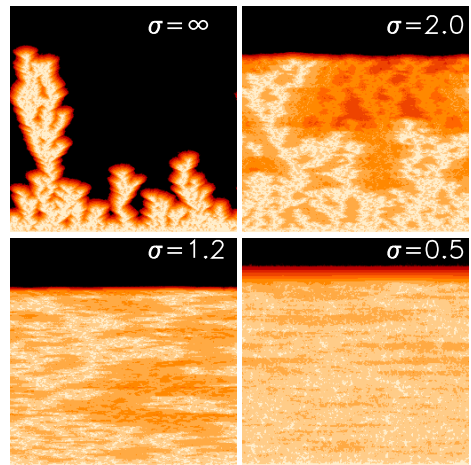}
\caption{
(Color online) Space-time plot of synchronization error
  $w_i(t)=|x_i(t)-y_i(t)|$ for two replicas of the CML (1-2) for $q=2$
  and $M=9$ with $\gamma$ slightly larger than $\gamma_c$ and
  different $\sigma$'s, see labels. The space (resp. time) axis
  correspond to the horizontal (resp. vertical) axis. The colors code
  the absolute value of the synchronization error in logarithmic
  scale. Colors go from intense white (yellow online) corresponding to
  $w_i(t) \sim 1$ to dark black associated to the fully synchronized
  state, i.e. $w_i(t)=0$.}
\label{fig1}
\end{figure}

To characterize the ST it is useful to introduce the synchronization
error $w_i(t)=|x_i(t)-y_i(t)|$.  Its spatio-temporal evolution is
reported in Fig.~\ref{fig1} for various values of $\sigma$ just above
the synchronization transition ($\gamma \gtrsim \gamma_c$). In the
short range limit $\sigma \to \infty$, percolating structures, typical
of ordinary DP, are clearly observable. As the range of the
interaction increases (i.e. $\sigma$ decreases) the spatial structures
tend to be smoothed out. Finally, for $\sigma\lesssim 0.5$, the spatial
structures are no more detectable.

To be more quantitative, we consider the spatial average of the
synchronization error $\rho_\gamma(t)=\sum_i w_i(t)/L$ which is the
natural order parameter: it vanishes, at sufficiently long times,
whenever a complete synchronization is achieved, i.e.,
$\rho^\ast_\gamma= \lim_{t\to \infty} \rho_\gamma(t)=0$ for
$\gamma>\gamma_c$; whilst it remains finite at any time in the
unsynchronized state, i.e. for $\gamma<\gamma_c$.  In terms of
$\rho_\gamma(t)$, we can now define the critical exponents $\delta$,
$\beta$ and $z$ that will be function of $\sigma$: $\delta$ rules
the temporal scaling of the order parameter at the critical point
$\gamma = \gamma_c$ i.e. $\rho_{\gamma_c}(t) \sim t^{-\delta}$;
$\beta$ controls the way the asymptotic value of the order parameter approaches
zero, $\rho^\ast_\gamma \sim
(\gamma_c-\gamma)^\beta $ for $\gamma \le \gamma_c$; and finally the
dynamical exponent $z$ can be defined in terms of the finite-size
scaling relation, valid at the critical point:
\begin{equation}
\rho_{\gamma_c}(t) \sim L^{-\delta z} f(t/L^z)\,.
\label{finite_size}
\end{equation}
\begin{figure}[t!]
\includegraphics[draft=false,clip=true,width=.495\textwidth]{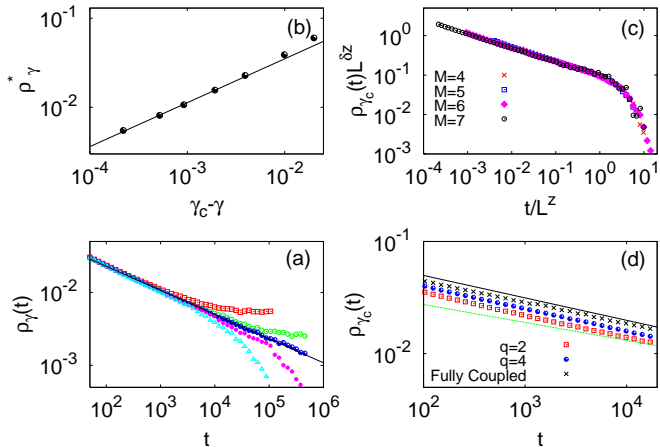}
\caption{(Color online) (a) Average synchronization error
  $\rho_\gamma$ vs time for the CML (\protect\ref{eq:cml}) with $q\!=\!4$
  and $M\!=\!8$ for $\sigma\!=\!1.4$, the curves refer to
  $\gamma\!=\!0.2977\,,$ $0.29789\,,\;0.29792\,, \;0.29797\,,\; 0.2981$
  (from top to bottom). Here $\gamma_c\!=\!0.29792(2)$. The straight line
  indicate the slope $\delta\!=\!0.33(2)$ as obtained by a best fit. (b)
  Asymptotic synchronization error $\rho_\gamma^*$ vs
  $\gamma_c-\gamma$ for the same parameters as in (a), the straight
  line indicates the slope $\beta\!=\!0.49(5)$. (c) Finite-size scaling
  $\rho_{\gamma_c}(t) L^{\delta z}$ vs $t/L^z$ from which we estimate
  $z\!=\!1.05(5)$, symbols refer to different sizes $L\!=\!2\, q^{M}$ (here
  $q\!=\!4$) and $M$ as in the legend.  (d) The synchronization error
  $\rho_{\gamma_c}$ vs time at criticality for $\sigma\!=\!2$ using $q\!=\!2$
  with $M\!=\!16$ and $q\!=\!4$ with $M\!=\!8$ together with the results for the
  fully coupled model for $\sigma_{fc}\!=\!1+\sigma\!=\!3$ and size
  $L\!=\!8,193$. The top straight-line displays the slope $\delta\!=\!0.2$ as
  estimated by a best fit; the bottom straight-line shows the (short
  range) DP value $\delta\!=\!0.15945$.  Data are obtained by averaging
  $\rho_\gamma(t)$ over many ($\sim 100$) different initial
  conditions.}
\label{fig2}
\end{figure}

These exponents, which fully characterize the transition (ruling the
divergence of, e.g., the spatial and temporal correlation lengths at
the critical point), have been measured for different values of
$\sigma$, by means of the following procedure (see
Fig.~\ref{fig2}a,b and c for an illustration of the method and to
appreciate the quality of the results).  The critical coupling
$\gamma_c$ is estimated by examining the temporal behavior of 
$\rho_\gamma(t)$, averaged over many random initial
conditions, for different values of $\gamma$.  In particular,
$\gamma_c$ is determined (typically with an absolute error within the
range $[2 \times 10^{-5}: 2 \times 10^{-4}]\,$) as the point at which
a scaling behavior for $\rho_\gamma(t)$ emerges (see
Fig.~\ref{fig2}a). With this method $\delta$ is
directly computed by means of a best fit procedure on
$\rho_{\gamma_c}(t)$.  Once $\gamma_c$ is known, $\beta$
is obtained by measuring the saturation values $\rho^\ast_\gamma$ in the
desynchronized regime (i.e. for $\gamma<\gamma_c$) for sufficiently
long chains, see Fig.~\ref{fig2}b. Finally, performing several
simulations with different sizes at the critical point
$\gamma=\gamma_c$, the exponent $z$ is estimated through a data
collapse based on Eq.~(\ref{finite_size}).  Although this
procedure works satisfactorily (Fig.~\ref{fig2}c) it is affected
by large errors due to the unavoidable subjectivity in judging the
collapse.

We remark that, as typical of strongly nonlinear
maps~\cite{baroni,cencini_torcini}, $\gamma_c$ cannot be identified by
the condition of vanishing transverse Lyapunov exponent and, moreover,
it depends not only on $\sigma$ but also on $q$.  However, the
critical exponents, apart from finite-size effects, are not influenced
by $q$ as one can judge from Fig.~\ref{fig2}d, where the
synchronization error at the critical point is shown for $q=2,4$ and
$\sigma=2$. In the figure we also display the result for the fully
coupled model with the exponent ruling the interaction range set to
$\sigma_{fc}=3$. The data confirm that $\sigma_{fc}=1+\sigma$ as
previously discussed.

\begin{figure}[t!]
\includegraphics[draft=false,clip=true,height=0.29\textwidth]{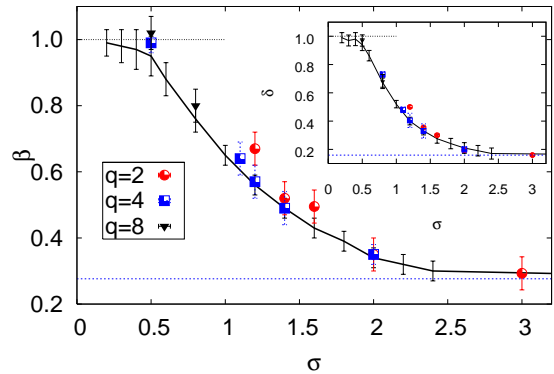}
\caption{(Color online) Critical exponent $\beta$ vs $\sigma$ obtained
  from best fits using different lengths and basis $q=2,4,8$. The
  black thick curve represents the data of HH \protect\cite{HH98}, the
  bottom straight line is the DP value, while the top one represents
  the mean field prediction by HH. CML with sizes from
  $L=2^{16}\approx 6\times 10^4$ up to $L=2\times 8^7\approx
  4\times10^6$ maps have been used.  Inset: Critical exponent $\delta$
  vs $\sigma$ symbols are as in the main panel.}
\label{fig3}
\end{figure}

\begin{table}[b!]
\begin{center}
\begin{tabular}{|c|c|c|c|c|c|c|c|} 
\hline $\sigma\;[q]$ & 3.0 [2] & 2.0 [4]& 1.4 [4]& 1.2 [4] & 0.8 [8] &
0.5 [8] \\ \hline $\Delta$ & 0.93(6) & -0.13(9) & -0.04(12)& -0.03(10)
& -0.03(23)&0.04(13)\\ \hline
\end{tabular}
\end{center}
\vspace{-0.5truecm}

\caption{Scaling relation (\ref{hyper}) for various values of
$\sigma$, for each measurement the corresponding basis $q$ is
reported. The measured $\Delta$ are compatible with zero in the range
$[\sigma_m:\sigma_c]$. See text for details.}
\label{tab1}
\end{table}

In Fig.~\ref{fig3}, we display the exponents $\beta$ and $\delta$ as a
function of $\sigma$, and compare them with the results of HH obtained
for stochastic models of L\`evy-flight epidemic spreading~\cite{HH98}.
The agreement between the two sets of data is rather good.  Moreover,
by increasing $q$ and so the maximal affordable spatial length $L$, we
observe that the exponents approach those of HH. This leads us to
conclude that discrepancies are mainly due to finite-size effects. The
results for $z$ (not reported) also confirm the overall agreement. Due
to the strong finite-size effects and to computational limitations it
was not possible to analyze systems with $\sigma < 0.5$. For anomalous
DP, it has been shown that in the interaction range
$[\sigma_m:\sigma_c]$ the following scaling relation
holds~\cite{janssen,HH98}
\begin{equation}
\Delta = 1 -\sigma +(1 - 2 \delta) z \equiv 0\,.
\label{hyper}
\end{equation}
As shown in Table \ref{tab1} the relation (\ref{hyper}) is fulfilled
within the errors, strengthening the parallel between ST in
spatio-temporal chaotic systems with power-law interactions and
anomalous DP.  It is worth stressing that it is absolutely nontrivial
that the deterministic system here investigated was in quantitative
agreement with the stochastic one studied in HH. The fact that the
critical exponents are the same suggests that the
universality class (actually the family of universality classes
labeled by $\sigma$) is the same for both models.

The agreement of the critical exponents with those of Ref.~\cite{HH98}
(shown in Fig.~\ref{fig3}) represents a strong indication that the
synchronization dynamics associated to model
(\ref{eq:cml}-\ref{eq:smartcoupling}) should admit the same stochastic
field description proposed for anomalous directed
percolation~\cite{janssen,HH98}:
\begin{equation}
\partial_t n=D_N\nabla^2 n+D_A\nabla^\sigma n +\tau n-\lambda n^2+g(n) \xi
\label{eq:langevin}
\end{equation}
where, in the present context, $n$ represents the coarse grained
synchronization error obtained by averaging $w_i(t)$ over a suitable
space-time cell.  The parameters entering Eq.~(\ref{eq:langevin}) are
the standard and anomalous diffusion coefficients $D_N$ and $D_A$,
respectively; $\tau$ that measures the distance from the critical
point (i.e. from the synchronization threshold $\gamma_c$) and the
amplitude of the nonlinear term $\lambda$. The fractional derivative
$\nabla^\sigma$ is non-local and can be defined through its action in
Fourier space $\nabla^\sigma \!\to -|k|^\sigma$. Finally, $\xi$ is a
zero-average $\delta$-correlated (in space and time) Gaussian noise
field with unit variance and $g^2(n)\! \propto\! n$. For systems with
short range interactions, Eq.~(\ref{eq:langevin}) with $D_A\!=\!0$ (i.e.
the field description of ordinary DP) was heuristically derived in
\cite{GLPT03} from a simple stochastic model, mimicking the
behavior of $w_i(t)$ for Bernoulli maps.  That derivation can be
straightforwardly extended to the present case giving
Eq.~(\ref{eq:langevin}) as a macroscopic description of the
dynamics. Indeed the only difference comes from the spatial coupling
that can be handled as in \cite{chaos} (where the synchronization
of long range coupled chaotic oscillators was considered) leading to
the fractional derivative term.  However, a detailed derivation goes
beyond the scopes of the present Letter.

In conclusion, we have provided strong numerical evidence that the
synchronization transition of two coupled replicas of a chain of
discontinuous maps with long range interactions is characterized by a
continuum of universality classes labeled by the exponent $\sigma$,
which sets the interaction range. Moreover, the critical exponents
agree with those reported for anomalous directed
percolation~\cite{HH98}.  Preliminary results indicate that also for
continuous maps the critical exponents depends on $\sigma$ and that
they do not belong to the anomalous DP universality class.  By analogy
with the present results, we conjecture that the critical properties
of the synchronization transition for continuous maps should be
reproducible in terms of a multiplicative noise Langevin equation with
long-range interactions (i.e. an equation similar to
(\ref{eq:langevin}) but with $g(n) \propto n$~\cite{munoz_review}).
To the best of our knowledge this equation has not been yet subject of
theoretical or numerical analysis.  It will be a definitely worth task
to undertake such study in the future

We are grateful to F.~Cecconi and F.~Ginelli for useful discussions,
remarks and for a careful reading of the manuscript.
M.C. acknowledges partial support from the PRIN2005 ``Statistical
mechanics of complex systems'' by MIUR and MPIPKS for computational
resources.


\begin{thebibliography}{99}
\bibitem{nota1} The results here presented are expected to extend to
  continuous maps with sufficiently large first derivatives, see
  Refs.~\cite{baroni,PA01,GLPT03,cencini_torcini} for a detailed
  discussion of this issue.
\end{thebibliography}
\end{document}